\documentclass[prd,showpacs,preprint,eqsecnum,amsmath,amssymb,nofootinbib]{revtex4-2}

\usepackage{amsfonts}
\usepackage{graphicx, epsfig, amsmath}
\usepackage{amssymb,amsmath, mathrsfs}
\usepackage{amssymb, graphics, setspace}
\usepackage{dcolumn}
\usepackage{bm}
\usepackage{color}
\usepackage{epstopdf}
\usepackage{hyperref}	
\newcommand{\la}{\langle}
\newcommand{\ra}{\rangle}
\newcommand{\w}{\omega}

\newcommand{\be}{\begin{equation}}
\newcommand{\ee}{\end{equation}}
\newcommand{\bea}{\begin{eqnarray}}
\newcommand{\eea}{\end{eqnarray}}
\newcommand{\bes}{\begin{subequations}}
\newcommand{\ees}{\end{subequations}}

\def\pbh{ {\cal H}_b^- }
\def\fbh{ {\cal H}_b^+ }
\def\pch{ {\cal H}_c^- }
\def\fch{ {\cal H}_c^+ }
\def\pni{ {\cal I}^- }

\def\kb{\kappa_b}
\def\kc{\kappa_c}
\def\kn{\kappa_N}

\begin{document}

\begin{titlepage}
\vfill

\begin{center}
\baselineskip=16pt
{\Large\bf Horizons and Correlation Functions in $2D$ Schwarzschild-de Sitter Spacetime}

\vskip 0.8cm

{\bf   Paul R. Anderson$^{(a)}$
and Jennie Traschen$^{(b)}$}

{$^{(a)}$
{Department of Physics, Wake Forest University, Winston-Salem, NC 27109, USA}\\
{$^{(b)}$ Department of Physics, University of Massachusetts, Amherst, MA 01003, USA }}\\

\vspace{0.3cm}
\vskip 4pt

\vskip 0.1 in Email: \texttt{anderson@wfu.edu, traschen@umass.edu}

\vskip 1.mm

\end{center}
\vskip 0.3in
\par
\begin{center}
{\bf Abstract}
\end{center}
\begin{quote}
Two-dimensional Schwarzschild-de Sitter is a convenient spacetime in which to
 study the effects of horizons on quantum fields since the spacetime
contains two horizons, and the wave equation for a massless minimally coupled scalar field can be solved exactly.
The two-point correlation function of a massless scalar is
 computed in the Unruh state. It is found that the field
 correlations grow linearly in terms of a particular time coordinate that is good in the future development of the past horizons, and that
  the rate of growth is equal to the sum of the black hole plus cosmological
 surface gravities.  This time dependence results from additive
 contributions of each horizon component of the past Cauchy surface that is used to define the state.
 The state becomes the Bunch-Davies vacuum in the cosmological far field limit.  The two point function for the field velocities is also analyzed and
 a peak is found when one point is between the black hole and cosmological horizons and one point is outside the future cosmological horizon.

\end{quote}
\end{titlepage}

\section{Introduction}

Studies of quantum fields in black hole and cosmological spacetimes have found important and fascinating
phenomena, including quantum mechanical production of particles related to black hole horizons \cite{Hawking:1974sw}  and cosmological horizons \cite{Gibbons:1977mu},
 as well as the prediction that
quantum fluctuations during inflation lead to scale invariant density fluctuations on large scales due to the cosmological
horizon \cite{Guth:1982ec}\cite{Starobinsky:1982ee}. Properties of particle production, quantum
fluctuations, and correlation functions have been the subject of extensive research  in cosmological and black hole spacetimes.  In this paper we compute the two-point correlation function of
 a massless minimally coupled scalar field in  Schwarzschild-de Sitter spacetime (SdS).  This spacetime contains both black hole and cosmological
 horizons, so the geometry is a compromise between the attraction of the black hole and the rapid expansion
 due to the positive
  cosmological constant $\Lambda$ . There are two unequal temperatures, two horizon areas that contribute
 to the total entropy, and the far-field is an expanding FRW cosmology. Hence many aspects of SdS do not
 fit neatly into a classic thermodynamic picture, allowing for interesting generalizations in semiclassical gravity of various features found for black holes in asymptotically flat spacetimes, and the possibility that new features will be discovered.

 An important simplification that we make is to work in two-dimensions rather than four.  This allows the two-point function to be computed analytically which makes the study of its properties more tractable. Since $2D$ gravity is not dynamical,
 our calculations  should be viewed as results about quantum fields in $2D$ curved spacetime, which are interesting in their own right. In addition, $2D$ calculations
are useful as a simplification of the $S$-wave sector in $4D$. We build on a body of previous studies. The Unruh vacuum for SdS in $2D$ was introduced
in  \cite{Markovic:1991ua} and here we further study its properties. (See also~\cite{Tadaki:1990a,Tadaki:1990b}).
Properties of one choice of the Unruh vacuum in $2D$ de Sitter with no black hole  were studied in \cite{Aalsma:2019rpt}.
 Features of having two temperatures in $2D$ SdS were examined in \cite{Choudhury:2004}, and thermal effects in $2D$
de Sitter were studied in \cite{Davies:2020rlx}. Reference \cite{Corley:1997ef} constructed a lattice black hole
model for a $2D$ Schwarzschild black hole to analyze entropy on a more fundamental level. By focusing on
the $S$-wave sector, that was
extended to  $4D$ to look at loss of information inside a black hole \cite{Lowe:2015eba}.  Note that $2D$ gravity
 coupled to
a scalar field is dynamical,  including  the Jackiw-Teitleboim  and the Callan–Giddings–Harvey–Strominger models.
For example,  Ref.\cite{Aalsma:2021bit} explores features of
information in  de Sitter in Jackiw-Teitleboim  gravity.  However in this paper we will take
the metric to be fixed and focus on the properties of a minimally coupled scalar field.

 The properties of a quantum field in a fixed classical spacetime depend on the choice of the quantum
 state.  We will use the Unruh state for SdS constructed in \cite{Markovic:1991ua}, which
 generalizes that of Schwarzschild spacetime
 \cite{Unruh:1976db}, in which particle states are
defined with respect to geodesic (or Kruskal) coordinates on the black hole and cosmological horizons. Working in  $ 2D$ SdS, Ref.\cite{Markovic:1991ua} showed that  the stress-energy tensor has the same behavior in this state
as is found at late times in the case of an SdS black hole that forms from the collapse of a massive shell.
Here we build upon that work and
calculate the symmetric two-point correlation function  $G^{(1)} $ and the velocity correlation function that can be derived from it.  In spite of the substantial simplifications that occur in $2D$, nontrivial quantum effects still occur.

Looking at correlations between points on slices of constant time $T$, we find that  $G^{(1)} $ increases linearly in $T$
with the rate determined by the sum of the surface gravities of the black hole and cosmological horizons,
\be\label{timedep}
G^{(1)}( T, r_ 1 , T, r_ 2 ) = {1\over 2\pi  } (\kb + \kc )T +  (r_1 , r_2 \ {\rm dependent \ term})
\ee
Here $T$ is a Killing coordinate that is well-behaved throughout the spacetime, as discussed below.
To  better understand the role of the horizons in leading to the
 result (\ref{timedep}), the two-point functions in $2D$ Minkowski, Schwarzschild, and deSitter, spacetimes along with a Bose-Einstein condensate (BEC) analog black hole, are computed.  We argue that the form of the
time dependence in (\ref{timedep}) arises from the relation between the geodesic and the Killing
coordinates on a  past horizon $J$, and that there is a contribution of $\kappa_J T$ from each horizon that is a
component of the past Cauchy surface. We also compute $G^{(1)} $ in the case that a Schwarzschild black hole
is formed by collapse of a null shell,
 when the quantum field is in the natural {\it in} vacuum state. We find, at late times, the same linear growth in time as for the eternal black hole examples mentioned above, illustrating that the
behavior (\ref{timedep}) can also occur in the absence of past horizons.

 For a BEC analog black hole the density-density correlation function is a measurable quantity.  It can be computed by taking various combinations of derivatives of a quantity that includes the two-point function for a massless minimally coupled scalar field in the analog spacetime~\cite{Balbinot}.
If scattering of the phonons is ignored then there is a negative peak that is predicted to occur in the density-density correlation function when one point is inside and one point is outside the future horizon of the analog black hole~\cite{Balbinot,Carusotto,Anderson:2013ux}.  This peak has been observed experimentally~\cite{Steinhauer-1,Steinhauer-2}.  For a realistic black hole such a peak might be expected to occur in correlation functions that involve derivatives of the two-point function and perhaps in the two-point function itself.  However, it is of less interest because it cannot be measured without passing through the event horizon of the black hole.
We note that a theoretical model has also been constructed for an analog model of particle production from a
 cosmological horizon \cite{Fedichev:2003id}.

To follow up on the structure found in the BEC system for $2D$ SdS, we computed the two-point correlation function for the field ``velocities" $\partial_T \phi$, and found that
peaks occur depending on where the points are located relative to the horizons.
For example, a peak  occurs when one point is in the static patch and one point is in the cosmological region.

 This paper is organized as follows:
The coordinate systems that we use for SdS are summarized in
 Section \ref{sectwo}. In Section \ref{secthree}, the Unruh vacuum for SdS is defined and
 the two-point function $G^{(1)}$ for $\phi$ is computed.
In Section \ref{other-spacetimes},  the time dependence of $G^{(1)}$ for points
at equal times is computed  for Minkowski and several black hole spacetimes.  The velocity correlation function in SdS is computed and studied in Sec.~\ref{velsec}. In Sec.\ref{cosmo}
the relationship between the Unruh state and the Bunch-Davies state in the far field limit is investigated  A brief summary and conclusions are given in Sec.\ref{concl}.

\section{Some useful coordinate systems for SdS }\label{sectwo}

There are several sets of coordinates that we find useful for SdS.  Here we review those coordinates and the relations between them.

First in the static region between the black hole and cosmological horizons one can define the usual static coordinates for which the metric is
\be\label{staticmetric}
ds^2   =  -f dt^2  +{1 \over f} dr^2  \ = f \left( - dt^2 +dr_*^2\right)
\ee
where the tortoise  coordinate $r_*$ is defined by $ dr_*  =  dr /f $, and
\be\label{fofr}
f(r)=1-\frac{2M}{r}-  H^2 r^2   = \ - \frac{H^2}{r} (r-r_c ) (r-r_b )(r +r_c +r_b )
\ee
Here $r_c > r_b $ are the locations of the cosmological and black hole horizons respectively, and
 $\Lambda = 1/ l_c^2= H^2$. The two parameterizations are related by
 \be
 M = {r_br_c(r_b+r_c)\over 2 (r_b^2+r_c^2+r_br_c)},\qquad H^2 = {1\over r_b^2+r_c^2+r_b r_c}  \label{M-H2}
\ee
For SdS  $\Lambda$ is positive and there is a maximal area for the black hole which occurs when
 $r_c =r_b = 1/( \sqrt{3} H )$, corresponding to a maximal mass in the 4D spacetime of $M_{max} = 1/( 3\sqrt{3} H )$ and
a surface gravity of zero.
 The ingoing and outgoing radial null coordinates are $u= t-r_* \ , \quad v=t+r_*$. An explicit expression  for $r_*$
 is given in (\ref{rstar}).

The wave equation for a massless minimally coupled scalar field in the metric (\ref{staticmetric}) is
\be\label{wave}
\left( \partial_t^2 -\partial_{r_*}^2 \right) \phi = 4 \partial_u \partial_v \phi =0  \;.
\ee
Hence the solutions are freely propagating waves which can be written as the sum of an arbitrary function of $u$
 and an arbitrary function of $v$.
Though the static coordinates $t$ and $r_*$ are useful for finding solutions to the wave equation, they
 are singular on the horizons, so we use  Kruskal coordinates which are geodesic
 on the horizons to define the  particle states. The Kruskal coordinates $U_b, V_c $ are geodesic
 on the past black hole $\pbh$ and cosmological horizon  $\pch$ respectively, and we choose
  $U_b =0$  on $\fbh$ and $V_c =0$ on $\fch$.
The transformation between the Kruskal coordinates and the static null coordinates $u$ and $v$ depends on the
region of the spacetime. One finds
\bea\label{UVuv}
 U_b  &= &  \frac{1}{\kappa_b} e^{-\kappa_b u} \   , \ \   r<r_b \ , \quad {\rm and} \quad
 U_b  = - \frac{1}{\kappa_b} e^{-\kappa_b u} \ , \ \  r>r_b \\  \nonumber
V_c & = &- \frac{1}{\kappa_c} e^{-\kappa_c v}  \  ,\ \  r<r_c \, \quad {\rm and}
 \quad V_c =  \frac{1}{\kappa_c} e^{-\kappa_c v}  \ \  , \ \   r>r_c
  \eea
where $\kappa_b$ and $\kappa_c$ are the magnitudes of the surface gravity of the black hole and cosmological
 horizons. A second set of Kruskal coordinates can likewise be defined. $V_b, U_c $ are geodesic on the future horizons  $\fbh$ and  $\fch$, and are related to the first set by
\be\label{krukru}
\kc V_c = - (\kb V_b )^{-\kc /\kb } \ , \quad \kb U_b = - (\kc U_c )^{-\kb /\kc } \;.
\ee

A third set of coordinates that we will make use of in looking at surfaces of constant time, since they are
also good across the future black hole and cosmological horizons,
 were found in \cite{Gregory:2017sor}\cite{Gregory:2018ghc}. Let
 \be\label{coord1}
 T = t + h(r ) \ ,\ \quad {\rm where} \quad { dh \over dr}  = {j \over f} \ , \quad  \quad j (r) =-\gamma r +{\beta \over r^2 } \;,
 \ee
and
 \be\label{gammabeta}
\gamma = {r_c^2 +r_b^2 \over r_c^3 - r_b^3 } \ , \quad \beta = {r_c^2 r_b^2 (r_b + r_c ) \over r_c^3 - r_b^3 } \;.
\ee
Then the 2D SdS metric becomes\footnote{The $4D$ SdS metric in these coordinates has an additional
$r^2 d\Omega ^2$ term.}
\be\label{regularmetric}
ds^2   =  -f(r)  dT^2 + 2j(r) dr dT +{1-j^2 \over f} dr^2 \;.
\ee
The constants $\gamma$ and $\beta$ have been chosen such that  $j (r_b ) = 1$ and $j (r_c ) =-1$, which
ensures that $T$ interpolates between the ingoing null coordinate $v$  at
the future black hole horizon and the outgoing null coordinate $u$ at the future cosmological horizon. These are
ingoing and outgoing Eddington-Finklestein coordinates respectively.
\footnote{$T$ also serves as a good coordinate  for a slowly rolling inflaton field and
perturbatively evolving metric in $4D$ \cite{Gregory:2018ghc}.}
The metric (\ref{regularmetric})
is well behaved in the static and cosmological regions and has the Eddington-Finklestein form on both the future black hole and future cosmological horizons. The coordinate $T$ stays timelike and $r$ stays spacelike beyond the cosmological horizon.

The integrations for $r_*$ and $T$  each contain an arbitrary constant  which we choose such that
\be\label{Tbc}
 T=u \  on \ \fch \  ,\quad \quad {\rm and} \quad  T=v  \ on \ \fbh  \;.
 \ee
 We find
\bea\label{Tcoord}
T=  t  &+ & h(r)  \\ \nonumber
= t  &+ &
{1\over 2\kb} log  {| r-r_b | \over r_c - r_b  } +{1\over 2\kc} log  {| r-r_c | \over r_c -r_b  } +
{1\over 2}\left( {r_c \over  r_b \kb }   -{1\over \kn}\right) log   { r +  r_c + r_b \over  r_c + 2r_b}  \\ \nonumber
 &- & {r_b r_c\over 2(  r_c - r_b)} log {r^2 \over r_c r_b}  +{r_c \over 4 r_b \kb } log { r_c + 2 r_b\over 2 r_c +  r_b }  \;,
\eea
and
\bea\label{rstar}
 r_* (r) &  =& {1\over 2\kb} log  {| r-r_b | \over r_c - r_b }- {1\over 2\kc} log  {| r-r_c | \over r_c  -r_b}
 + {1\over 2\kn} log   {| r+r_c + r_b  | \over r_c + 2r_b } \\ \nonumber
 & - & {r_c \over 4r_b \kb }log{2r_c + r_b \over r_c +2r_b}  -{r_b r_c \over 2 ( r_c - r_b )}log{r_b\over r_c} \;,
\eea
where at a horizon $r = r_h$ we set $2\kappa _h = |f' (r_h )| $,
 so that all of the surface gravities denote positive quantities. Here $r_N =-(r_b + r_c )$ and
 refers to the negative root of $f(r)$. For future use, the surface gravities are given by
  \bea\label{surfgravs}
  \kb &=& {H^2 \over 2 r_b } (r_c -r_b ) (r_c + 2r_b )  \;, \\ \nonumber
    \kc &=& {H^2 \over 2 r_c } (r_c -r_b ) (2r_c + r_b )  \;, \\ \nonumber
      \kn &=& {H^2 \over 2(r_c +  r_b ) } (2r_c + r_b ) (r_c + 2r_b ) \;.
      \eea
While there are several quantities of geometrical interest, including
 $H , M,  r_b , r_c , \kb $ and $\kappa_c$, it is important to recall that SdS is only a two-parameter family of solutions.
 We will generally take the two independent parameters to be $H$ and $r_b$.
 .

 To relate the Kruskal coordinates to $T$ and $r$, note that
 $u= T-( h(r) +r_* ) $ and $v= T- h(r) + r_*  $. Substitution  into~\eqref{krukru} gives
 \be\label{vutr}
  V_{c} = {1\over \kc}  e^{-\kc T}  \tilde{V}_c   \ ,\  \quad and \quad
 U_{b} ={1\over \kb} e^{-\kb T}  \tilde{U}_b \;,
\ee
where $\tilde{U}_b$ and $\tilde{V}_c$ only depend on $r$ and are given by
\bea\label{krutor}
 \tilde{V}_c &=&    { r-r_c  \over r_c -r_b}
\left({ r+r_c +r_b \over  r_c + 2 r_b}  \right)^{r_b / 2r_c } \left( { r_b\over r} \right)^{H^2 r_b (2 r_c +r_b )/2 }  \;,
\\ \nonumber
 \tilde{U}_b &=&  -  {  ( r-r_b ) \over  r_c - r_b}
\left({ r+r_b +r_c \over  2r_c + r_b}  \right)^{r_c / 2r_b } \left( { r_c\over r} \right)^{ H^2 r_c ( r_c +2 r_b ) /2} \;.
\eea
These expressions can be used in any region, that is, there is no longer a set of different cases.  When working
through the algebra, one finds that the different signs in the relations for the transformations between
$U_b , V_c$ and $u,v$ given in (\ref{UVuv})  are compensated for by the
different cases encoded in the absolute value signs in $h$ and $r_*$.

\section{Two-point function for SdS in the Unruh state }\label{secthree}

The generalized SdS Unruh vacuum \cite{Markovic:1991ua}  defines particle states
 on the past Cauchy surface $\pbh \cup \pch$
 with respect to modes that are well behaved on the horizons. Black hole and cosmological particle production
 in the Unruh vacuum was computed in \cite{Qiu:2019qgp}.
 It has been shown~\cite{Markovic:1991ua}  that this state gives the same behavior for the stress-energy tensor
as is found at late times in the case of an SdS black hole that forms from the collapse of a massive shell in 2D.  Since we are ultimately interested in black holes that form from collapse in the early universe, the Unruh state is the natural one for us to consider.

 Using the Kruskal coordinates, the boundary conditions on the past modes are
\bes \bea\label{unruhpart}
 p^b_\w &=& \frac{1}{\sqrt{4 \pi \w}} e^{-i \w U_b} \quad on\ \pbh \;, \\
&=&0 \quad  \quad \quad on\ \pch \\
     p^c_\w &=& \frac{1}{\sqrt{4 \pi \w}} e^{-i \w V_c} \quad on\ \pch \;, \\
     &=&0 \quad\quad\quad on\ \pbh \;.
      \eea \ees
The enormous simplification in $2D$ is that these functions are solutions to the wave equation throughout
the spacetime.

The expansion for the field in terms of  Kruskal modes is
\be \phi = \int_0^\infty d \w \left[ a^b_\w p^b_\w + a^c_\w p^c_\w +
 a^{b\dagger}_\w p^{b*}_\w + a^{c\dagger}_\w p^{c*}_\w \right] \;, \label{phi-exp} \ee
where the $a^b_\omega $ and $a^c_\omega$ annihilate the Unruh vacuum
\be\label{anni}
a^b _\omega  |0 \ra = a^c _\omega  |0 \ra = 0 \;.
\ee
The creation and annihilation operators are normalized as $[ a^h_\omega ,   a^{h'\dagger}_{\omega'}]
 = \delta(\w-\w') \delta^{h, h'} $, where $ h,h' = b,c$.
The symmetric Green's function is
\be\label{symmgreen}
 G^{(1)}(x,x') =  \la 0| \phi(x) \phi(x') + \phi(x') \phi(x)|0 \ra \;.
  \ee
Substituting~\eqref{phi-exp} into \eqref{symmgreen} then gives
\bea\label{green2d}
  G(x,x') &=&  \int_0^\infty d\w [ p^b_\w(x) p^{b \, *}_\w(x') + p^b_\w(x') p^{b \, *}_\w(x) + p^c_\w(x) p^{c \, *}_\w(x') + p^c_\w(x') p^{c \, *}_\w(x)] \nonumber \\
   &=& \frac{1}{2 \pi} \int_{\omega_0}^\infty \frac{d \w}{\w} \left\{ \cos[\w (U_b - U_b^{'})] + \cos[\w(V_c - V_c^{'})] \right\} \nonumber \\
    &=& - \frac{1}{2 \pi} \left\{ ci[\omega_0 |U_b - U_b^{'}|] + ci[\omega_0 |V_c - V_c^{'}|] \right\} \;.
    \eea
with $ci$ the cosine integral function and $\omega_0$ an arbitrary infrared cutoff. We have dropped the
superscript $(1)$ on $G$ since we will only be considering the symmetric Greens function.
   For small values of its argument the cosine integral can be approximated as $ci(z) = \gamma_E + \log z$ with $\gamma_E$, Euler's constant.  For the calculations done in this paper we assume that $T' = T$.  Then, for any fixed value of $\omega_0 > 0$ and for fixed values of $r$ and $r'$, the arguments of the cosine integral functions in~\eqref{green2d} will be small enough to use this approximation for large enough values of $T$.  This is the implicit assumption that is made in our analysis of the time dependence of the two-point function for SdS and also for the other cases studied where horizons exist.  For SdS one finds
\be\label{greenapprox}
 2\pi G(x,x') = -2 \gamma_E -  \log \omega_0^2 |\Delta U_b \Delta V_c | \;.
 \ee

\subsection{Correlations on equal time slices}
Next we analyze the behavior of the correlation function when the two points have the same
time coordinate $T$,  since this is a good time coordinate in the static and cosmological regions and on the future horizons.
The case when one point is on each future horizon is particularly simple, and provides guidance for the general case.

\subsubsection{Horizon to Horizon Correlations}
Let $x_1$ be on $\fbh$ and $x_2$ on $\fch$, then making use of
(\ref{krukru}),  and~\eqref{Tbc}
gives
\bes \bea
\Delta U_b & = & -{1\over \kb} (\kc U_{c2} )^{-\kb /\kc} =  \ -{1\over \kb} e^{-\kb T} \;, \\
\Delta V_c & = &   {1\over \kc} (\kb V_{b1} )^{-\kc /\kb} =  \ {1\over \kc} e^{-\kc T } \;.
\eea \label{dudvone} \ees
Substituting into~\eqref{greenapprox} gives
\be\label{greenshh}
2\pi G(x_1 , x_2 ) =  (\kb + \kc) T  - \log \left( { \omega_0^2 \over \kb \kc   } \right)  -2\gamma_E \;.
\ee
Hence correlations between the horizons grow linearly in $T$ with the growth rate set by the sum of the surface gravities.

\subsubsection{General case}

Next we analyze $G$ for two points that have the same $T$ coordinate but otherwise can be anywhere.
 Substituting $U_b$ and $V_c$ given in (\ref{vutr}) and (\ref{krutor}) into (\ref{greenapprox}) yields
 \be\label{greengen}
2\pi G(x_1 , x_2 ) = T(\kb +\kc )
 - log\left(  {\omega_0^2 \over \kb\kc } |  \Delta \tilde{U}_b  \Delta \tilde{V}_c |   \right) -2\gamma_E \;.
\ee
Hence we learn that quite generally for points at the same time $T$, $G$ increases in time like $T(\kb +\kc ) $.
This is one of our main results.

\begin{figure}[htbp]
	\includegraphics[width=70mm]{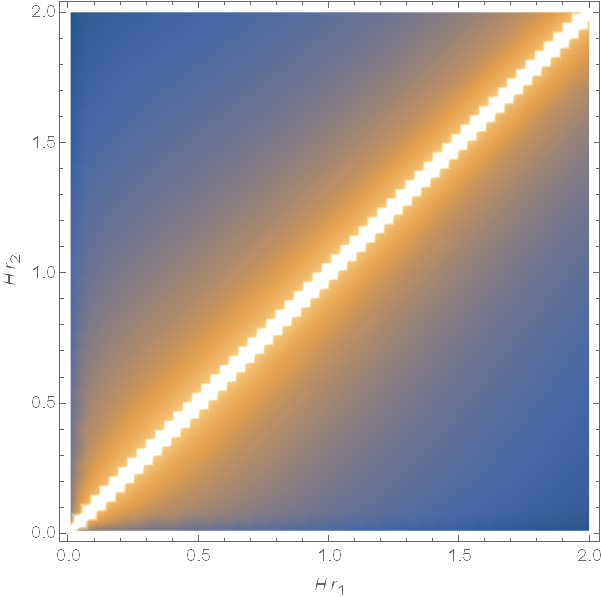}
\hspace{0.3cm} \includegraphics[width=11.5mm]{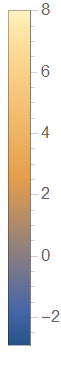}
	\caption{Density plot of $\bar{G}(r_1,r_2)$ for $ H  r_b = 0.2$.  The horizontal axis is $H r_1$ and
the vertical axis is $H r_2$.    \label{fig:G}}
\end{figure}

The spatial dependence of $G(x_1,x_2)$ is
\be \label{gbar}
2 \pi \bar{G}(r_1,r_2) \equiv  - \log( |  \Delta \tilde{U}_b  \Delta \tilde{V}_c |   )
\ee
where $ \tilde{U}_b ,  \tilde{V}_c$ are given in \eqref{krutor}. The correlation function~\eqref{gbar}
is not translation invariant, so its value depends on the location of each of the points. One way of plotting this is shown in Fig.~\ref{fig:G}.
Note that as usual the Green's function diverges on the
coincidence line $r_1 = r_2$, and then decreases monotonically away from this line in both directions.
Thus the most interesting feature of $G(x_1 , x_2 )$ is its time dependence. We will see below that
there is more structure in the two-point function
for the time derivative of the field $\partial_T \phi$.

\section{$\kappa T$ dependence in other 2D spacetimes}
\label{other-spacetimes}

Having found the simple linear time dependence of $G$, and that the sum of the surface gravities determines the
rate of change of correlations, it is useful to compare these results to the corresponding correlation functions in other spacetimes. In particular we choose Minkowski,
Schwarzschild, and de Sitter spacetimes,  which `deconstruct' the SdS case  along with
a BEC analog black hole metric.  All of these have static regions bounded in the past by either
past null infinity or a past horizon and bounded in the future by future null infinity or a static region.

\subsection{Minkowski Spacetime}
In $1+1$ dimensions we take
\be\label{flat}
ds^2 = -dt^2 + dr^2 = \ - du dv \;,
\ee
where $u=t-r ,\ v=t+r$ as before, and all the coordinates range between plus and minus infinity.
An early time Cauchy surface can be taken to be the union of the two branches of past null infinity, and
the modes are
 \be\label{flatmodes}
 p_\omega^{in} = {1\over \sqrt{ 4\pi \omega } } e^{-i\omega (t-r ) } \ \ , \quad
  p_\omega^{out} = {1\over \sqrt{ 4\pi \omega } } e^{-i\omega (t+r ) }  \;.
  \ee
 The Green's function is derived by the same steps as in (\ref{green2d}) and (\ref{greenapprox})
  with $p^b$ replaced by $p^{in}$ and  $p^c$ replaced by $p^{out}$, which gives
\bes \bea\
2\pi G_M (x_1 , x_2 ) &= &  - log|\omega_0^2 \Delta u \Delta v | -2 \gamma_E   \\
&= & -  log|\omega_0 \Delta r |^2 -2 \gamma_E \ , \quad\quad t_1 =t_2
\eea \label{gflat} \ees
At equal times the Green's function is independent of $t$.

\subsection{Schwarzschild Spacetime}\label{schw}
Setting $H=0$ in (\ref{regularmetric}) gives the Schwarzschild metric in $(T,r)$ coordinates,
\be\label{schwarz}
ds^2 = - \left(1- {r_b \over r } \right) dT^2 + 2{r_b^2 \over r^2 } drdT + \left(1 + { r_b  \over r }\right) \left(1 + \frac{r_b^2}{r^2} \right)  dr^2 \;,
\ee
which is good on the black hole future horizon, where  $T$ becomes the ingoing null coordinate $v $.
The surface gravity is $\kb =1/(2 r_b )$.
 The past Cauchy surface is now the past black hole horizon plus
past null infinity $\pni$. The particle states on $\pbh$ are defined by the modes $p_\omega^b (U_ b) $ as in SdS,
and states on $\pni$ are defined with respect to the ingoing modes in static coordinates $p_\omega^\infty (v) $.
Then the calculation of $G_{S} $ follows the same steps as in SdS that lead to (\ref{green2d}), but
 with $p_\omega^\infty (v) = e^{-i\omega v}/\sqrt{ 4\pi \omega}  $ replacing $p_\omega^c (V_c ) $, which gives
 the Schwarzschild Greens function in the Unruh vacuum,
\bea\label{gs}
2\pi G_{S,U} (x_1 , x_2 ) & = &  - \log|\omega_0^2 \Delta U_b \Delta v | -2 \gamma_E \\ \nonumber
 & = &  - \log {\omega_0^2 \over \kb}| \Delta \left(e^{\kb (-T + r_* +h)  } \right) \Delta (T+ r_* -h ) | -2 \gamma_E  \;.
\eea
In the second line  we have  used
$U_b =-e^{-\kb u} / \kb$, $u=T-h -r_*$, and $v=T-h + r_*$.
This is distinct from the Hartle-Hawking vacuum, which is explicitly constructed to be thermal.

Setting $T_1 =T_2 =T$ we see that $G_{S,U}$ increases in time like $\kappa_b T$,
\be\label{gstwo}
2\pi G_{S, U} (x_1 , x_2 ) = \kappa_b T - 2\gamma_E - \log\left| {\omega_0^2 \over \kb }
( e^{\kb (h+r_{*} )_2 }  - e^{\kb (h +r_{*} )_1 } ) (r_{*2} - h_2 - r_{*1}+ h_1  ) \right| \;,
\ee
where $h(r) +r_*  = r + r_b  ( 2 \log {| r-r_b | \over r_b } -   \log {r\over r_b } ) $ and $
 r_* -h (r)   = r + r_b \log {r \over r_b}$.
There is only one surface gravity factor in $G_{S,U}$ since there is only one past horizon.

We have used the $(T,r)$ coordinates rather than the static Schwarzschild $(t , r )$ coordinates,
since here $G (T, r )$ is well behaved when one point is on the future horizon.
For example, setting $r_1 =r_b$ and $r_2 =r \gg r_b$
in (\ref{gstwo}) gives the finite and simple relation
\be
2\pi G_S (r_b , r ) \simeq \kb ( T -r)   - {3\over 2} \log( \omega_0 r ) - \frac{1}{2} \log(\omega_0 r_b)
\ee
A question that is relevant for Schwarzschild spacetime is what happens for the Hartle-Hawking state?
 The difference between the Hartle-Hawking state and the Unruh state comes from the modes that originate at past null infinity.  In the
Hartle-Hawking state these are positive frequency with respect to the Kruskal coordinate $V$.
Outside the past and future horizons $V_b = \kappa^{-1} e^{\kappa v}$.  Thus instead of $p^\infty_\w(v)$ one has $p^\infty_\w(V_b) = e^{-i \w V}/\sqrt{4 \pi \w}$.  The resulting two point function is
\bea\label{gsHH}
2\pi G_{S,U} (x_1 , x_2 ) & = &  - \log|\omega_0^2 \Delta U_b \Delta V_b | -2 \gamma_E \\ \nonumber
 & = &  - log {\omega_0^2 \over \kb^2}| \Delta \left( e^{\kb (-T + r_* +h)  } \right) \Delta \left( e^{\kb (T + r_* -h )} \right) | -2 \gamma_E  \;.
\eea
Setting $T_1=T_2=T$ gives
\be\label{gsHHtwo}
2\pi G_{S, U} (x_1 , x_2 ) = - 2\gamma_E - \log\left| {\omega_0^2 \over \kb^2 }
( e^{\kb (h+r_{*} )_2 }  - e^{\kb (h +r_{*} )_1 } )  ( e^{\kb (-h+r_{*} )_2 }  - e^{\kb (-h +r_{*} )_1 } )\right| \;.
\ee
So there is no $T$ dependence for the Hartle-Hawking state.

 \subsection{de Sitter Spacetime}\label{2dds}
 Setting $M=0$ in (\ref{regularmetric}) gives the 2D de Sitter metric
  in coordinates that are good on the cosmological horizon,
 \be\label{dsone}
ds^2 = -  (1- H^2 r^2 )dT^2 -2Hr dT dr + dr^2 \;.
\ee
These are related to the usual static coordinates  with metric
\be
ds^2 = - (1-H^2 r^2) dt^2 + \frac{dr^2}{1-H^2 r^2} \  =  (1- H^2 r^2 )\left( -dt^2 +  dr_*^2 \right)
 \;,  \label{dS-static}
\ee
by the coordinate transformations  $ T = t + \frac{1}{2 H} \log |1 - H^2 r^2|$ and
$ r_* = \frac{1}{2 H} \log | \frac{1+ H r}{1-H r} |$.

To get to the Bunch-Davies state one further transforms to the cosmological coordinates. Here $T$ is
equal to cosmological proper time and letting $r = \rho e^{H T}= -\frac{\rho}{H \eta} $
 gives
\be ds^2 = -d T^2 + a^2 d \rho^2 = a^2 (-d \eta^2 + d \rho^2)  \;, \label{dS-cosmolog} \ee
with the scale factor
\be a = e^{H T} = \frac{-1}{H \eta} \;. \label{a-dS} \ee
In 2D dS space one can take the coordinate $r$ to range over the entire real line with horizons
  at $r = \pm H^{-1}$, or $ \eta = \pm \rho$, and a Cauchy surface is
 the union of the two branches of the past cosmological horizon. Alternatively, reflecting boundary conditions
 can be put at $r=0$ as is done in \cite{Aalsma:2021bit}. We make the first choice as it meshes nicely with
 the construction with a black hole.

 The general solution to the wave equation for the massless scalar field
  can be written as a function of $(\eta + \rho )$ and a function
 of $(\eta - \rho )$. The Bunch-Davies state is defined by the positive frequency modes
 \be\label{bdmodes}
 p_\omega^{in} = {1\over \sqrt{ 4\pi \omega } } e^{-i\omega (\eta - \rho ) } \ \ , \quad
  p_\omega^{out} = {1\over \sqrt{ 4\pi \omega } } e^{-i\omega (\eta + \rho ) } \;.
  \ee
These modes are the same as Kruskal modes $e^{-i\omega U} ,\ e^{-i\omega V} $, which can be checked as follows:  The Kruskal coordinates are
\be U = - {1\over H} e^{-H u} \quad {\rm and} \quad  V = \mp {1\over H} e^{-H v} \;, \label{Kruskal-dS-1} \ee
with  $u =t-r_* $, and $v=t+r_*$ as before.
The Cauchy surface is at $U=-\infty$ and $V=-\infty$, the future horizons being at $U=0$ and $V=0$.
Using the coordinate transformations gives
\be  U = \eta - \rho\;, \qquad {\rm and} \qquad V = \eta + \rho  \;. \label{Kruskal-dS-2}  \ee
Hence the Bunch-Davies state (\ref{bdmodes}) is the same as the Unruh state.

The Green's function has the same form as (\ref{green2d}) with $p^b$ replaced by $p^{out}$ and
$p^c$ replaced by $p^{in}$.  At equal times $T_1=T_2=T$ one finds
 \be
2\pi G_{dS} (x_1 , x_2 )
=  2 HT  -  log( \omega_0 \Delta r )^2 - 2\gamma_E \;.   \label{gds}
\ee
In de Sitter $H=\kc$,  so the time dependence of $G$ has the same form as was found in SdS with
$\kb$ replaced by $\kc$ from the ``other" past cosmological horizon.

This would seem to be the natural way to split the points in the radial direction when one considers de Sitter space in
the coordinates (\ref{dsone}), which are natural from the point of view of taking a limit of $SdS$ with very small mass.
 On the other hand, in cosmological coordinates the natural way to split the points in the radial direction is to have $T_1=T_2=T$ and $\rho_1 - \rho_2 = \Delta \rho$.  Then using $r = e^{H T} \rho $ one finds that the linear growth in $T$ in~\eqref{gds} disappears. In retrospect, it is not surprising that with all the symmetries of de Sitter space,
  there are different ``natural" answers depending on the choice of observers.

%

\subsection{BEC analog black hole spacetime}

In~\cite{Balbinot,Carusotto,Anderson:2013ux} a simple model of a Bose-Einstein condensate analog black hole was investigated.  It was assumed that the BEC is effectively confined to one dimension
with a steady flow and a sound speed $c(x)$ that varies with position along the flow. In some experiments~\cite{Steinhauer-1,Steinhauer-2} and subsequent models~\cite{waterfall-1,waterfall-2} the flow speed $v_0$ also varies with position.
For a flow in the $-\hat{x}$ direction an analog black hole exists if there is a subsonic region where $v_0 < c$,
and then to the left of this there is a supersonic region where $v_0 > c$.  The sonic horizon occurs where
$v_0 = c$. Theoretical models show that in the hydrodynamic or long wavelength approximation, the phonons satisfy a wave equation that has the same form as the wave equation for a massless minimally coupled scalar field in a certain spacetime.  After dimensional reduction to 1+1 dimensions, the resulting mode equation is a 2D wave equation for a scalar field with an effective potential that depends on the sound speed and flow velocity profiles.  So long as the sound speed and flow velocity approach constant values in the limits $x \to \pm \infty$, the spacetime associated with this wave equation has the same type of structure as Schwarzschild spacetime with the exception that there are no singularities inside the horizons.

In the original calculation of the BEC density-density correlation function~\cite{Balbinot}
 the effective potential in the mode equation for the phonons was ignored.  This resulted in a mode equation that is the same as that for a massless minimally coupled scalar field in the analog spacetime.  This is the approximation that we shall use here.  The metric for the analog spacetime is
\be ds^2 =  \frac{c(x)^2 - v_0(x)^2}{c(x)} (- dt^2 + d x^2_{*})     \;,  \label{analog-metric} \ee
the tortoise coordinate is
\be x_* = \int^x d y \frac{c(y)}{c(y)^2 - v_0(y)^2} \;, \label{BEC-xstar-def} \ee
and the surface gravity of the analog black hole is
\be \kappa_b = \left. \left( \frac{d c}{dx} - \frac{d v_0}{d x} \right)\right|_{\rm hor}  \;. \ee
One can then define the null coordinates $u = t - x_*$ and $v = t + x_*$.

For the BEC analog black hole experiments described in~\cite{Steinhauer-1,Steinhauer-2}, the density was measured throughout the condensate at a particular laboratory time $T$ and used to compute the density-density correlation function.  The laboratory time is given by the expression
\be T = t + \int^x dy \frac{v_0(y)}{c(y)^2 - v_0(y)^2} \;. \label{BEC-T-def} \ee
Using~\eqref{BEC-T-def} and~\eqref{BEC-xstar-def} in~\eqref{analog-metric} one finds
\be d s^2 = - \frac{(c^2-v_0^2)}{c} d T^2 + 2 \frac{v_0}{c} dT dx + \frac{1}{c} dx^2 \;, \ee
which is of the same form as~\eqref{regularmetric} with $f = c^{-1} (c^2-v_0^2)$, $j = \frac{v_0}{c}$, and $h = \int^x dy \frac{v_0(y)}{c(y)^2-v_0(y)^2}$.
One finds that the null coordinates are
\bea u &=& T - \int^x \frac{dy}{c(y) - v_0(y)} \;, \nonumber \\
     v &=& T + \int^x \frac{dy}{c(y) + v_0(y)} \;. \eea

As for Schwarzschild spacetime, in this BEC analog black hole spacetime the Cauchy surface can be taken to be a
union of past null infinity and the past horizon, with
particle states on $\pbh$ defined by the modes $p_\w^b(U_b)$,  and the particle states on $\pni$ are defined with respect to the ingoing modes in static coordinates $p^\infty_\w(v)$.
Hence the two-point function is
\be 2 \pi G_{{\rm BEC}, U}(T_1,x_1;T_2,x_2) = - 2 \gamma_E - \log |\w_0^2 \Delta U_b \Delta v|   \;. \ee
This has the same form,~\eqref{gs}, as it does for Schwarzschild spacetime.
Setting $T_1 = T_2 = T$ one finds
\bea  & & 2 \pi G_{{\rm BEC}, U}(T,x_1;T,x_2) = - 2 \gamma_E + \kappa_b T - \log \left(\frac{\w_0}{\kappa_b} \right) - \log \left| \w_0 \int_{x_1}^{x_2} \frac{dx}{c(x)+v_0(x)} \right| \nonumber \\  && \nonumber \\
&&  - \log \left| \exp\left(\kappa_b \int^{x_2} \frac{dy}{c(y) - v_0(y)}\right) \pm \exp\left(\kappa_b \int^{x_1} \frac{dy}{c(y) - v_0(y)}\right)\right| \;, \eea
where the plus sign occurs if one point is inside and the other point is outside the future horizon and the minus sign occurs if both points are on the same side of the horizon.
As in the other 2D black hole cases there is a linear growth in $T$.

 \subsection{Generalizing: Killing horizons and time dependence}

In the preceding examples the horizons are Killing horizons, and particle states are defined with respect to geodesic coordinates on a past Cauchy surface.  On the horizon(s), this is a Kruskal coordinate.
Each past horizon contributes a time dependence to $G$ of $\kappa_J \tau$,
 where $\tau$ is a good coordinate. We expect that this behavior is quite general when the Cauchy surface
 contains Killing horizons, which can be seen as follows. An
 affine parameter $\lambda$  for the null geodesic generators of a
 Killing horizon is related on the horizon to the Killing parameter $s$ by
 \be\label{affine}
 \lambda \sim e^{\pm \kappa s} \;,
 \ee
 (see {\it e.g.} \cite{wald:1984}\cite{carter:1986})
 where the plus and minus options allow for our convention that $\kappa$ is the magnitude of the surface gravity.
 In the examples worked out here, $\lambda$ is one of the $U$ or $V$ coordinates.
  Since there is no scattering for the massless minimally coupled scalar field in 2D, the modes have the same form away from
  the horizon as they do on it.  Extending the relation (\ref{affine}) off the past horizon yields $s = \tau + R'  $ where $\tau$ is a good timelike coordinate  on the past horizon
 and $R'$ is a spacelike coordinate. The coordinate $T$ used in the cases considered above is
 a good timelike coordinate  on the future horizon(s),  and a further coordinate transformation can be made to give
 $s = T + R$, with $R$ a different spacelike coordinate.

Hence when the initial Cauchy surface contains one or two components $J$ that are horizons, since
well-behaved particle states are defined with respect to $\lambda_J$  and since there is no scattering, in the bulk of the spacetime
 those modes oscillate in  $\lambda_J \sim e^{\pm \kappa_J T }   e^{\pm \kappa_J R } $.
  It is this exponential relation between
 $\lambda_J$ and $\kappa_J T$ that gives rise to a contribution of  $\kappa_J T$ to $G$. Further,  we see that
 the additivity
 $G = \sum_J \kappa_J T +...$ arises when the Cauchy surface contains multiple components.

\subsection{Collapsing null shell spacetime}

 Black holes formed from gravitational collapse,
  in which case $\pbh$ does not exist, are not included in the general argument above. So we turn
  to a dynamical shell collapse example, and show that at late times the same time dependence of
  the two point function is recovered.

The shell starts at past null infinity $\mathscr{I}^{-}$ and moves along the trajectory $v = v_0$ for some constant $v_0$.  Inside the shell the spacetime is flat.  In 4D the mode functions must be regular at $r = 0$. Equivalent boundary conditions can be obtained by putting a static perfectly reflecting mirror at $r = 0$.  Outside the shell the metric is that for Schwarzschild spacetime.  In 4D, given that the radial coordinate $r$ is related to the area of a two sphere, it is useful to have $r$ be continuous across the shell and $t$ discontinuous. Alternatively one can have $v$ be continuous and $u$ discontinuous. Letting the subscript $F$ refer to coordinates in the interior flat region
 and imposing continuity of $v$ gives  $u = t - r_*$ and $u_F = t_F - r$ is~\cite{Massar:1996tx,Fabbri:2005mw}
\be
u = u_F - \frac{1}{\kappa} \log \left( \kappa (v_H-u_F)\right)  \;,
\label{u-u_F}
\ee
with
\be
v_H \equiv v_0 - 4M  \;.
\label{vH-def}
\ee
Inverting this gives~\cite{mirror-bh}
\be
u_F = v_H - \frac{1}{\kappa}\, W\left[ e^{\kappa(v_H - u)} \right]  \;,
\label{u_F-u}
\ee
with $W$ the Lambert $W$ function.

Inside the null shell the modes for the {\it in} vacuum state are
\be
p^{\rm in}_\w = \frac{1}{\sqrt{4 \pi \w}} (e^{-i \w v} - e^{-i \w u_F})  \;. \label{pin-null-shell}  \ee
Because the mode equation has the general solution $p = f(u) + g(v)$ outside the shell and the mode functions are continuous across the shell,~\eqref{pin-null-shell} is the form of the solutions outside the null shell as well with $u_F(u)$ given by~\eqref{u_F-u}.

Since $p^{\rm in}_\w$ vanishes at $\w = 0$, the two point function is infrared finite rather than infrared divergent. Substituting~\eqref{pin-null-shell} into~\eqref{symmgreen} gives
\bea G(x,x') &=& \int_0^\infty d \w \; [p^{\rm in}_\w(x)\,  p^{{\rm in} \; *}_\w(x') +  p^{\rm in}_\w(x') \,  p^{{\rm in} \; *}_\w(x) ] \nonumber \\
   &=& \lim_{\w0 \to 0} \frac{1}{2 \pi} \int_{\w_0}^\infty \frac{d \w }{\w}\; \left\{\cos[\w(v-v')] - \cos[\w(v-u_F(u'))] - \cos[\w(v'-u_F(u))] \right. \nonumber \\
     & & \left. \qquad + \cos[\w(u_F(u)-u_F(u'))] \right\} \nonumber \\
   &=& \lim_{\w0 \to 0} \frac{1}{2 \pi}   \left\{- {\rm ci}[\w_0(v-v')] + {\rm ci}[\w_0(v-u_F(u'))] + {\rm ci}[\w_0(v'-u_F(u))] \right. \nonumber \\
   & & \left. \qquad - {\rm ci}[\w_0(u_F(u)-u_F(u'))] \right\}  \nonumber \\
   &=&  \frac{1}{2 \pi} \log \left[\frac{(v-u_F(u'))(v'-u_F(u))}{(v-v')(u_F(u)-u_F(u'))} \right]
       \;. \label{G-null-shell} \eea

Using $v = T - h(r) + r_*$ and $u = T - h(r) - r_*$, one finds that at a late time $T$ with  $r$ fixed,
\be u_F(u) \to  v_H - \frac{1}{\kappa} e^{\kappa (v_H-u)}  = v_H - \frac{1}{\kappa} e^{\kappa v_H} e^{-\kappa T} e^{\kappa (h(r) + r_*)}, \nonumber \\
         \ee
Then
\bea  u_F(u) - u_F(u') &\to& \frac{e^{\kappa v_H}}{\kappa} (-e^{-\kappa T} e^{\kappa (h(r) + r_{*})} + e^{-\kappa T'} e^{\kappa (h(r') + r_*')} \nonumber \\
  v - u_F(u') &\to& v - v_H = T - h(r) + r_{*} - v_H \nonumber \\
   v'-u_F(u) &\to&  v' - v_H = T' - h(r') + r'_{*} - v_H \;. \eea
Substituting into~\eqref{G-null-shell} and setting $T' = T$ gives in the late time limit
\bea 2 \pi G(x,x') &\to& \kappa T + \log (\kappa^2 T^2)  - \kappa v_H - \log [\kappa |r_* - h(r) - r_*' + h(r')|] \nonumber \\
 & & - \log [ |-e^{\kappa(h(r) + r_*)} + e^{\kappa (h(r')+r'_{*})}|]  \eea
To leading order this is the same growth in time $\kappa T$
 as that for static Schwarzschild spacetime when the field is in the Unruh state.  However, here there is a subleading term that also grows in time.  Further, the growth only occurs in the late time limit.  For early times $u_F \approx u$ and there is no linear growth in $T$.

\vspace{0.3cm}

\section{Velocity correlation function}\label{velsec}

As discussed in the Introduction, a correlation peak has been measured in the density-density correlation function for a BEC analog black hole~\cite{Steinhauer-1,Steinhauer-2}.
This correlation function can be obtained by taking various combinations of derivatives with respect to the spacetime coordinates of a quantity closely related to the two-point function.  When scattering is neglected the correlation peak occurs when one point is inside and the other point is outside the future horizon.  If scattering of the mode functions is taken into account then other correlation peaks are also found.

In this section the two-point function for the field velocities $\partial_T \phi$ is computed as an example of a correlation function that depends on two derivatives of the two-point function, one at each spacetime point.
For comparison we first record the $2D$ Minkowski result. Using the freely falling particle states and coordinates
 defined in (\ref{flat}), (\ref{flatmodes}) gives
\be\label{flatvv}
 2\pi \la \partial_{t_1} \phi (x_1 ) \partial_{t_2}\phi (x_2 ) +  \partial_{t_2} \phi (x_2 ) \partial_{t_1}\phi (x_1 )  \ra = -{1\over (u_2 -u_1 )^2 }
 -{1\over (v_2 -v_1 )^2 } \;.
 \ee
The correlation function is negative definite, goes to zero at null and spacelike infinity,
and at equal times $t_1 =t_2$ reduces to $-2/R^2$ with
$R=|x_2 -x_1 |$.  The only peaks are infinite ones that occur when the points come together or when both points are on the same null surface.

Returning to SdS, consider $\phi = F(U) +G(V) $, where
 in this section we set  $U\equiv U_b$ and $V\equiv V_c$  for ease in reading.
Then
\be\label{dot}
\partial_T \phi = -\kb U F'(U) - \kc VG'(V) \;.
\ee
Hence
\bea\label{dphidphi}
2\pi g (x_1 , x_2 ) & \equiv &  2\pi \la \partial_{T_1} \phi (x_1 ) \partial_{T_2}\phi (x_2 ) +  \partial_{T_2} \phi (x_2 ) \partial_{T_1}\phi (x_1 ) \ra \\ \nonumber
 & =& \lim_{\w_0\to 0} \left\{\kb^2 U_1 U_2 {\partial\over \partial U_1 }{\partial\over \partial U_2 }
\int_{x_0}^\infty  {dx \over x} \cos x
+\kc^2 V_1 V_2 {\partial\over \partial V_1 }{\partial\over \partial V_2 }
\int_{y_0}^\infty {dy \over y} \cos y \right\} \;,
\eea
where $x_0 = \omega_0 |U_2 - U_1|$ and $y_0 = \omega_0 |V_2 - V_1|$.  The limit $\w_0 \to 0$ can
be taken because after taking the derivatives there is no infrared divergence.  The result is
\bea\label{dphidphi2}
2\pi g (x_1 , x_2 ) &= & - \kb^2 {U_1 U_2 \over (U_2 -U_1 )^2 }
- \kc^2 {V_1 V_2 \over (V_2 -V_1 )^2 }  \\ \nonumber
&\equiv& 2\pi g_U (U_1 , U_2 ) + 2\pi g_V (V_1 , V_2)  \;.
\eea

 To examine these changes in more detail,
it is useful to write $g_U$ and $g_V$ in terms of the Schwarzschild null coordinates $u$ and $v$ using~\eqref{UVuv}.  The result is
\bes \bea  2 \pi g_U(u_1,u_2) &=&  - \frac{\kappa_b^2}{4 \sinh^2\left[\frac{\kappa_b}{2}(u_1-u_2) \right]} \;, \qquad r_1, r_2 < r_b \;\;{\rm or} \;\; r_1, r_2 > r_b  \;, \label{GUrrrb} \\
               &=&  \frac{\kappa_b^2}{4 \cosh^2 \left[\frac{\kappa_b}{2}(u_1-u_2) \right]} \;,  \qquad r_1 < r_b < r_2 \;\;{\rm or} \;\; r_2 < r_b < r_1 \;, \label{GUrrbr} \\
 2 \pi g_V(v_1,v_2) &=&    - \frac{\kappa_c^2}{4 \sinh^2\left[\frac{\kappa_c}{2}(v_1-v_2) \right]} \;, \qquad r_1, r_2 < r_c \;\;{\rm or} \;\; r_1, r_2 > r_c  \;, \label{GVrrrc} \\
               &=&  \frac{\kappa_c^2}{4 \cosh^2 \left[\frac{\kappa_c}{2}(v_1-v_2) \right]} \;,  \qquad r_1 < r_c < r_2 \;\;{\rm or} \;\; r_2 < r_c < r_1 \;. \label{GVrrcr}
\eea \label{GUGV}\ees

In the analysis that follows, we assume that $T_1=T_2=T$ and that $H$ is held fixed.  Then from~\eqref{M-H2} it is seen that
 the cosmological radius is given by
\be\label{rc}
H  r_c = {1\over 2 } \left( - H r_b + \sqrt{ 4 - 3H^2 r_b^2 }\right) \;.
\ee
The black hole has a maximum size, corresponding to a minimal size of the cosmological horizon, when
$r_b = r_c =1/(\sqrt{3} H)$.  In this extremal case $\kappa_b = \kappa_c = 0$ so there is no Hawking radiation and no Unruh state.

\begin{figure}[htbp]
	\includegraphics[width=100mm]{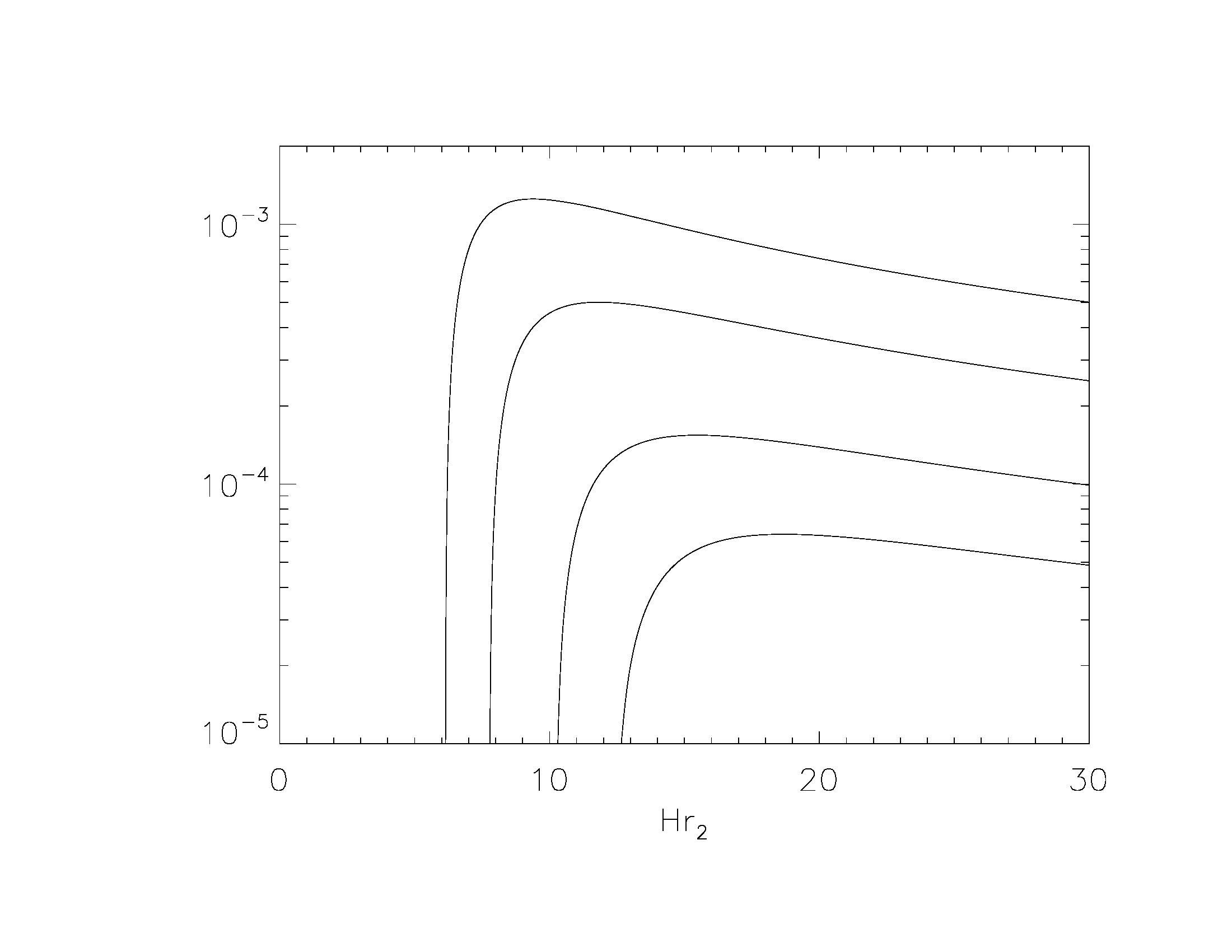}
	\caption{The velocity correlation function is plotted as a function of $H \, r_2$ for $ H\, r_b = 0.1$ and $H \,r_c \approx 0.946$.
The  curves from top to bottom correspond to $H \,r_1$ being fixed at $0.9 H r_c$, $0.95 H r_c$, $0.98 H r_c $, and $0.99 H r_c$ respectively.
Since the purpose of this figure is to illustrate the changes in size and location of the peaks as the value of $r_1$ changes, only the parts of the curves with $r_2 > r_1$ and $g(r_1,r_2) > 0$ are shown.   \label{fig:p7}}
\end{figure}

What happens when both points are on opposite sides of a horizon is
complicated because there can be comparable contributions from $g_U$ and $g_V$.
We focus here on the potentially observable peaks that occur when one point is in the static region and the other is in the cosmological region.
In this case, if $r_1$ is fixed in the static region there is always a correlation peak
 at some $r_2$ in the cosmological region.  To see this first note that
 for $r_b < r_2 < r_c$, $g(x_1,x_2) < 0$.  As $r_2$ increases $g_U$ remains negative but decreases.  However, $g_V$ goes through zero at $r_2 = r_c$ and is then positive for all larger values of $r_2$.  Now for $r_2 > r_c$,  both $g_U$ and $g_V$ decrease in magnitude as $r_2$ increases.  However, $g_U$ which is negative decreases in magnitude faster than $g_V$ because $\kappa_b > \kappa_c$.  Thus eventually $g(x_1,x_2)$ changes sign, becoming positive.  However, it also vanishes in the limit $r_2 \to \infty$.  Therefore $g(x_1,x_2)$ has a maximum at some value of $r_2 > r_c$.  This property is illustrated in Fig.~\ref{fig:p7}.  Note that as $r_1$ gets closer to the horizon, the peak gets smaller and is located at larger values of $r_2$.

\section{ Cosmological far field limit of the Unruh state, and relation to the Bunch-Davies vacuum}\label{cosmo}
Calculations of the power spectrum due to quantum fluctuations in models of inflation start with
the calculation of the two-point function for a scalar field in
the Bunch-Davies state.
 Ultimately we would like to know how this is changed by the presence of a black hole in $4D$,
or more realistically, by a population of black holes. In the context of our $2D$ calculations, we can examine
the relation between the Unruh state for SdS and the cosmological far field.
In Sec.~\ref{2dds} we showed by transforming coordinates that  the Unruh modes with $M=0$ are the same as the Bunch-Davies ones. In this section we find the leading effect of nonzero $M$ in the cosmological region.

To make contact with an asymptotically cosmological
 FRW form of the metric, the needed coordinate transformations from SdS
start with going to one of the ``McVittie" forms analyzed in \cite{Kaloper:2010ec}. Let
\be\label{coord2}
T= \tau + \xi (r)  \  , \quad {\rm where} \quad {d\xi \over dr } =
 {1\over f} \left( j + {Hr \over (1 - {2M \over r } )^{1/2} } \right) \;.
\ee
Then
\be\label{mcmetricone}
 ds^2
=  -f d\tau^2 -{2H r \over \left( 1 - {2M \over r } \right)^{1/2}  } dr d\tau +{ dr^2 \over 1-{2M\over r} } \;.
\ee
Asymptotically, $\tau$ is the cosmological proper time coordinate.
Next, transforming to an asymptotically comoving coordinate $\rho$
\be\label{coord3}
r =\left( 1+  {M\over 2 a \rho } \right)^2 a \rho \ \ , \quad  a =  e^{H\tau}  \;.
\ee
 yields the McVittie metric \cite{McVittie:1933zz} in two dimensions,
\be\label{mcmetrictwo}
ds^2 = a^2 \left[ - { \left( 1 - {M\over 2 a \rho } \right)^2 \over \left( 1 +  {M \over 2 a \rho }  \right)^2 } d\eta^2
+\left( 1 + {M\over 2 a \rho } \right)^4   d\rho^2  \right] \;.
\ee
Here  $\eta$ is asymptotically conformal time defined by $d\eta = d\tau/ a $, and
$  e^{H\tau}=  -{1\over H\eta} $,  with $ -\infty < \eta < 0$.

To find the form of the Kruskal modes  $p^b_\omega$ and $p^c_\omega$, given in (\ref{unruhpart}),
in the cosmological far field region, one needs the coordinate transformations when $r\gg r_c $, that is, when
\be\label{farthest}
a \, \rho = - \frac{\rho}{H \, \eta} \gg 1 \;.
\ee
In this limit
\be\label{farfield}
{  M \over 2a\rho } = - {MH\eta \over 2\rho }  \ll 1 \;,
 \ee
corresponding to late times $\eta\rightarrow 0$ or large distances $\rho\rightarrow \infty$, where
the metric (\ref{mcmetrictwo}) approaches the deSitter metric. Using
equations (\ref{coord1}) and (\ref{coord2}) and integrating the transformations in the limit
$Hr\gg 1 $ and $ {M\over r} \ll 1$ gives

\be\label{ttoeta}
t\simeq -{1\over H} \log (-H^2 \eta r ) +{M\over Hr} \ \quad and  \quad 2\kc r_* \simeq {2 \over H} {1+b \over r}
\ee
where we have switched to conformal time $\eta$, and where
 \be\label{bdef}
b = {H\over 2} \left( r_c -{2\over H} -{\kc r_b\over \kb} + {\kc (r_c +r_b )  \over \kn} \right) \;.
\ee
The definition of the constant $b$ is chosen so that $b=0$ when $M=0$. Combining the results gives
\be\label{uvfar}
t \pm r_* \simeq  -{1\over H} \log (-H^2 \eta r ) + {1\over H r} \left( M \pm {1+b \over \kc} \right) \;.
\ee
Lastly, transforming from $r$ to $\rho$ in (\ref{uvfar}) gives the far field coordinate relations
\bea\label{kruskalfar}
V_c & \simeq & {H\over \kc } (H\rho )^{\kc / H -1}\left[ \rho +\left( 1 + M \kappa_c +b \right)\eta \right] \;, \\ \nonumber
U_b  &\simeq &  {H\over \kb } (H\rho )^{\kb / H -1}\left[ -\rho + \left(\frac{\kappa_b}{\kappa_c} - M \kappa_b  + \frac{b \kappa_b}{\kappa_c} \right)\eta \right] \;.
\eea
When $M=0$, the formula for $V_c$ reduces to its value for de Sitter space in (4.16). The mode $\exp( -i\omega V_c )$
is ingoing and originates on $\pch$. The mode
 $\exp( -i\omega U_b )$ is outgoing and it describes the black hole radiation arising from $\pbh$.
One could scale $\rho$ and $\eta$ in such a way that $V_c$ and $U_b$
are proportional to $ \eta \pm \rho$, but one cannot eliminate
the overall factor of $\rho^{\kappa_c/H -1}$ in $V_c$ or the factor of $\rho^{\kappa_b/H -1}$ in $U_b$.  Therefore
there is a difference between the behaviors of the modes for the Unruh state when $M > 0$ and the modes
for the Bunch-Davies state when $M = 0$ even far from the black hole. One would expect that in $4D$
the difference from one in the exponent of $\rho$ will show up as a correction to the scale invariance of the
power spectrum. This interesting and potentially observable feature will be studied in subsequent work for
$4D$ SdS.

In Sec.~\ref{2dds}, it was shown that the linear growth in $T$ for de Sitter space disappeared if cosmological coordinates were used and the
points were split such that $T_1 = T_2 = T$ and $\rho_1 - \rho_2 = \Delta \rho \ne 0$.  For SdS the transformations are more complicated but one can see from~\eqref{kruskalfar} that in the far field region both $V_c$ and $U_b$ approach values that are independent of $\tau$ so long as $\rho > 0$ in the limit $\tau \to \infty$.  This means that there will be no linear growth in $\tau$ of the two-point function in the far field region when $\tau_1 = \tau_2 = \tau$ and $\rho_1 \ne \rho_2$.

With the blackhole, the SdS spacetime is inhomogeneous and the far field geometry is very different than
 the region close to the black hole, with the different regions being delineated by the horizons. Thus, we find it an
 an interesting result that there is linear growth in $T$ of the two-point function in the region between and near the cosmological horizon, even if there is no such growth in $\tau$ in the far field region. The most interesting result will be
to do the more complex calculation in $4D$ and find the impact on the CMBR.

\section{Concluding Remarks and Future Directions}\label{concl}

 The symmetric two-point function $G$ was computed analytically in the generalized Unruh vacuum.
Its  behavior was analyzed on surfaces of constant $T$, where $T$ is a  well-behaved time coordinate throughout the spacetime.
It was found that the time rate of change of $G$ is governed by the sum of the surface gravities
\be\label{dgdt-SdS}
{d \over dT }G (T, r_1 ; T, r_2 )  = \kappa_b + \kappa_c \;.
\ee

To determine how general this result is the two-point function was computed in several 2D spacetimes that contain past horizons.
   In each case it was found that
\be\label{dgdt}
{d \over dT }G (T, r_1 ; T, r_2 )  = \sum_J \kappa_J \;,
\ee
where the sum is over any past horizons used to define the vacuum on the initial Cauchy surface.
We gave an argument that this will be the case in general for
 2D spacetimes with Killing horizons.
In cases such as an asymptotically flat black hole
 where both a past horizon and past null infinity exists, the sum can be thought of as including
 a contribution from the latter surface since it has a surface gravity of zero. Further,
 the result  for $G$ is actually more general  also holding at late times for  a Schwarzschild black hole that forms from collapse of a null shell.

 The two point function of the field velocities ($\partial _T \phi$) was also computed.  When the two time coordinates are equal and the points are spatially separated, it
  is time-independent, but it exhibits some interesting spatial structure.
  In particular a peak was found when one point is in the static patch and one point is in the cosmological region.  This point separation makes it possible, at least in principle, to measure such a correlation peak if it persists in 4D.  A similar peak was found in the density-density correlation function for a BEC analog black hole~\cite{Balbinot}. Such a peak is also
  expected to occur for a Schwarzschild black hole, but an observer would have to pass through the event horizon to measure it.

It is important to ask if features of
the $2D$ results will be present in higher dimensions, and it is of considerable interest to compute the effect of
an early population of black holes has on CMBR anisotropies, extending the results of \cite{Prokopec:2010nm}.
The $2D$ calculation allows us to understand the stripped-down
problem with no scattering, and has turned out to be interesting in its own right. Wave propagation in $4D$
SdS is significantly more complicated.
Preliminary calculations show that 2D features may well persist for the contribution to the Greens function from the higher frequency modes, that is, modes with $\omega^2$ much larger than the maximum height of the scattering potential. Larger SdS black holes have a lower potential barrier, and so more of the modes propagate without significant scattering.  A detailed calculation is deferred for future work.

\acknowledgments
We thank David Kastor, Alessandro Fabbri, and Ian Moss for useful conversations.  This work came about as a result of conversations at the NORDITA
Gravitational Physics with Lambda Workshop 2018.
 We would like to thank NORDITA for their generous support.  This work was supported in part by the National Science Foundation under Grants No. PHY-1505875 and PHY-1912584 to Wake Forest University.
One of the plots was made using the WFU DEAC cluster; we thank the WFU Provost's Office and Information
Systems Department for their generous support.

\end{document}